\documentstyle[12pt]{article}

\def\be{\begin{equation}}
\def\bea{\begin{eqnarray}}

\def\ee{\end{equation}}
\def\eea{\end{eqnarray}}
\def\R{\rm {I\kern-.200em R}}
\def\C{\rm {I\kern-.520em C}}
\begin{document}

\begin{titlepage}
\hfill\vbox {\halign {&#\cr IPM-94-29 \cr hepth/9401158 \cr }}
\begin {center}

{\large { Vertex Operators of SL(2,R) Black Hole \\ and 2-d gravity }}
\\
\vskip 1cm{M. Alimohammadi$^{(a,b)}$ and F. Ardalan$^{(a,c)}$}\\
\vskip 2cm
$^a$ {\it Institute for Studies in Theoretical Physics and
Mathematics\\  P.O.Box 19395-5746, Tehran, Iran\\
 $^b$ Physics Department, Tehran University , North Karegar ,\\
 Tehran, Iran\\
 $^c$ Physics Department, Sharif University of Technology,\\
 P.O.Box 11365-9161, Tehran, Iran}
\end{center}
\vskip 1cm
\begin{abstract}
By boosting the vertex operators of Witten's $SL(2,R)/U(1)$ black hole, we show
that in the region
V they lead to the primary fields of $c=1$ matter coupled to gravity at nonzero
cosmological constant, while there is no such correspondence in the region I,
showing
that Witten's black hole corresponds to $2d$ gravity only in a certain
region
and in a specific limit. By using the free field representation , we
will show that the stress tensor of $SL(2,R)$ gauged by its nilpotent subgroup
is equivalent to that of the Liouville theory with zero cosmological constant.

\end{abstract}
\vskip 2cm
\vfill
\end{titlepage}
\noindet
\section{Introduction}
The study of the black hole background of string theory in the context of
gauged
Wess-Zumino-Witten (GWZW) models has attracted much attention in the last two
years
$^{[1-4]}$.One of the interesting features of these investigations is the
relation
between the Minkowski 2d black hole ,i.e. ,the $SL(2,R)$ theory with a $ U(1)$
subgroup gauged, and the
2d gravity, i.e., the Liouville theory coupled to $ c=1$ matter $^{[5-12]}$.

In Ref.[1] it was argued that as it is not possible to remove one of the
parameters
of the two dimensional black hole in favour of the Liouville field in all the
regions
(near infinity and near horizon of region I of Ref.[1] ),therefore the theory
can not be regarded as
a non-critical string theory of $ c=1$ matter coupled to gravity. In agreement
w
ith this result
Distler and Nelson$^{[6]}$ studied the BRST cohomology of the black hole and
found that there are more
discrete states in the black hole than in the Liouville
theory.Also by looking
at the behaviour of the states of the black hole near the horizon, Marcus and
Oz$^{[12]}$ found out
that there are only a few states that do not diverge near the horizon and
therefore are physical, and the $W_\infty$ states are not among them .

On the other hand using the free field realization of ${ SL(2,R)/ U(1)}$
and the true
BRST charge in the black hole it was shown that as far as the
energy-momentum tensor is concerened, the model is identical to 2d gravity
$^{[8,13]}$. Recently it has been argued that$^{[18]}$ there
are null states in the black hole which lead to even more discrete states
than in Ref.[6].
So the question of this relationship is open and requires further
investigation.
In this paper we will study the relation by a different
route,i.e.,by gauging the SL(2,R) by a nilpotent subgroup.

In Ref.[14] we studied the ${SL(2,R)/ E_1}$ theory ($E_1 $ is generated by a
nilpotent element $\sigma = \sigma_3 + i \sigma_2 $)
and found that at the quantum level an extra
coordinate is eliminated
and the resultant one-dimensional effective action is the Liouville action . To
understand
this elimination of a coordinate, we looked at ${SL(2,R)/ U_t (1)}$ , where
$U_t (1)$ is generated by $\sigma_3^t$ which
is a boosted $\sigma_3 (\sigma_3^t=e^{-{t \over2}\sigma_1} \sigma_3 e^{{t\over
2} \sigma_1})$.For finite values of $t$,this theory is the same as
$SL(2,R) / U(1) $
and is therefore two dimensional.But as $t \rightarrow \infty$ (where $
\sigma_3^t \rightarrow \sigma$ ) we found that
there is a remnant bosonic field which couples to the slowly varying Liouville
background in region V (Fig.1).(with coupling constant which vanishes at $t=
\infty$).This
indicated a relation between black hole and $ c=1$ matter coupled to gravity.

We will now continue
our investigation by studying the vertex operators of $SL(2,R) / U_t(1) $.
We will show that the primary fields in the regions V and III lead to the
vertex
operators of c=1+Liouville, but the fields of region I do not reduce to
c=1+Liouville.
In this way we will show that there is no correspondence between these two
theor
ies in all
the regions (in agreement with the result of Ref.1 in region I); but
otherwise
   the c=1+Liouville theory is only the limit of a black hole in the regions
III and V. We will also show that the eigenvalues of $ \sigma$ play the role
of cosmological constant in Liouville theory (as in Ref.[3]).

Finally we will construct the free field representation of the nilpotent
gauged WZW model of $SL(2,R)$
and show that the stress tensor of this model is the same as in the Liouville
theory.

\section {Operators in region I}

The primary fields of WZW model are defined via its matrix elements. In the
case
of $SL(2,R)$ gauged by $\sigma_3$, the vertex operator in the region I is
$^{[3]}$:
\be V_\lambda ^\omega = <\lambda ,\omega|g(y,\tau)|\lambda, -\omega> \ee
where $ \lambda$ defines the spin of  $SL(2,R)$  representation, $\omega$ is
the eigenvalue of $\sigma_3$ and $y$ and $\tau$ are defined by:
\be y=uv  ,\ \ \ \ \ \ \ \    e^{2\tau}= -{u \over v} \ee
where $u$ and $v$ are the parameters of $SL(2,R)$ group elements:
 \be g=  \left( \begin{array}{ll}  a&u\\-v&b  \end{array} \right),\ \ \ \ \
\ \ \ ab+uv=1 \ee
In region I, a suitable gauge condition is , $a-b=0$.

In Ref.[3] it was found that there are four different vertex operators in
region
I which have different behaviours near the horizon and infinity. Among
these fields,
one ,denoted $U_\lambda ^\omega$, can be naturally extended to the region III
and we will therefore
work with it first, as it can be extended to different regions of space-time.
Its explicit form is $^{[3]}$ :
\be  U_\lambda ^\omega= e^{-2i\omega\tau} F_\omega ^ \lambda
(y)\\=e^{-2i\omega\
tau} (-y)^{-i\omega}
B(\nu_+, \bar\nu_-),F(\nu _ +, \bar\nu_-,1-2 i\omega,y) \ee
where $B(\alpha, \beta )=\Gamma ( \alpha ) \Gamma ( \beta ) /
\Gamma(\alpha + \beta )}$, F is the hypergeometric function $_2F_1$ , and
\be \nu_\pm = {1\over 2} -i(\lambda \pm \omega) \ee
Now to find the relation between this
theory and
the c=1+Liouville theory we will look at the behaviour of vertex
operators of
$SL(2,R) / U_t(1) $. At finite values of $t$ , these operators are the same as
those of $SL(2,R) / U(1) $
, as $\sigma _3^t$ is conjugate to $\sigma_3$, and at $t=\infty$ they
will reduce
to $SL(2,R) / E(1) $, which we will show later to be equivalent to the vertex
operators of Liouville
theory. But the most interesting limit, is the case where $t$ is close to
infinity. As we will see, in this
limiting region a relation between the black hole and c=1+Liouville  theory
will
emerge. However, it is more convenient to work with the boosted group element
$g$ , in Eq.(1), rather than using the state corresponding to the $U_t(1)$. We
therefore have,
\be U_\lambda ^\omega (t)=<\lambda,\omega|g(y_{-t}, \tau_{-t})|\lambda
,-\omega>
   \ee
where
\be g_{-t}=e^{{t\over 2} \sigma _1}g e^{-{t \over 2 }\sigma _1}  \ee
Using Eqs.(7),
(3) and (2), and applying the gauge condition (which in
the boosted version is $a_{-t}-b_{-t}=0$), we arrive at the following form
for the parameters for large $t$:
$$ y_{-t}=u_{-t}v_{-t}=-{1\over4}(u-v)^2+(u+v)^2 e^{-2t}+O(e^{-4t})$$
\be \tau_{-t}={1\over2} \ln(-{u_{-t}\over v_{-t}}})={2(u+v)\over
u-v}e^{-t}+O(e^{-2t})\ee
Now to find the behaviour of Eq.(6) under $t\rightarrow \infty$, we must also
know the behaviour of the eigenvalues of $\sigma_3(\omega)$ under boosting .As
$\sigma_3|\lambda,\omega>=\omega|\lambda,\omega>$, therefore boosting
the subgroup $\sigma_3$ by $\sigma_3 \rightarrow \sigma_3^t=e^{-{t \over2}
\sigma_1}\sigma _3 e^{{t\over2} \sigma_1}$ must be followed by the
transformation
$|\lambda,\omega> \rightarrow |\lambda,\omega>_t=e^{-{t\over
2}\sigma_1}|\lambda,\omega>$,
such that the eigenvalues do not change under this similarity
transformation:
$$\sigma_3^t|\lambda,\omega>_t=\omega|\lambda,\omega>_t$$
If the eigenvalues of $\sigma=\sigma_3+i\sigma _2$ are denoted by $\chi$
$$\sigma|\lambda ,\chi >=\chi |\lambda ,\chi>$$
then as $\sigma^t_3 \stackrel {t \rightarrow \infty} \longrightarrow e^t
\sigma$ the states $|\lambda ,\omega>_t$ must tend
to $|\lambda ,\chi>$ as $t\rightarrow \infty$. As a result,
\be \omega=\chi e^t \ee
and
\be \nu_ \pm= {1\over 2}-i(\lambda \pm \chi e^t) \ee
Now there are two different possibilities which lead to
different limits for the above $\nu _\pm$, either
$\chi \neq 0$ which leads to $\nu _ \pm=\mp i\chi e^t \equiv \mp \nu$ , or
$ \chi \rightarrow 0$ in such away that $\omega=\chi e^t$ is finite.

To find the limit of the vertex operators for the first case, we will
first express the
hypergeometric function in terms of the associated Legendre functions of the
$2^
{nd}$
kind $Q_\mu ^\nu(z) $ $ ^{[15]}$:
$$Q_ \mu ^0 (z)={2^\mu \Gamma ^2(1+\mu) \over (z+1)^{\mu+1}
\Gamma(2+2\mu)}F(1+\mu , 1+\mu ,2(1+\mu) , {2\over 1+z})$$
The vertex operators then tend to,
$$U_\lambda ^\chi (t)\stackrel {t\rightarrow \infty}\longrightarrow e^{-4i\chi
{u+v \over u-v}}Q_ \nu ^0(1+{8\over(u-v)^2})$$ which when
$|\nu| \rightarrow \infty$ reduces to:
 \be U_\lambda ^\chi (t)\rightarrow e^{-4i\chi {u+v\over u-v}}
\sqrt {\pi \over 2}(w^2-1)^{-1\over 4}(w-\sqrt
{w^2-1})^{-{i\over 2}\chi e^t} \ee
where $w=1+{8\over (u-v)^2}$

In the second case, all the terms in $U_\lambda^\omega (t)$ are finite
and there are no simplifications in the $t \rightarrow \infty$ limit.

{}From these results it is apparent that there are no relations between the
vertex operators
of the region I  (at $t \rightarrow \infty$ limit) and the c=1+Liouville vertex
operators.
The same result is obtained if one consideres  the other vertex operators of
region I and no correspondence can be seen.Therefore we must seek this relation
in the other regions of the black hole.
\section {The Behaviour of Fields in the Region V as $t \rightarrow \infty$}

The natural form of the vertex operators in the region V are $^{[3]}$ :
\be W_\omega ^\lambda (y,\tau)=e^{-2i\omega\tau} y^{-i\omega} F(\nu_+ , \bar
\nu_- , 1 , 1-y) \ee
where $y=uv$ and $\tau= {1\over 2}\ln ({u/v})$ and the gauge condition is
$a+b=0$.
As this function has no singularity when crossing the singularity at $y=1$, it
c
an
be trivially continued to region III.

In the same way as discussed in the previous section, the corresponding vertex
operator
of ${SL(2,R)/ U_t(1)}$ in this region can be recovered from Eq.(12)
by simply transforming $y \rightarrow y_ {-t}$
, $\tau \rightarrow \tau _{-t}$ and taking the gauge condition
$a_{-t}+b_{-t} =0$
. Applying this gauge to the boosted parameters, it
can be shown that:
$$ y_{-t}=u_{-t}v_{-t}=e^ {2\varphi} e^{2t} + (2B e^{2 \varphi}-x^2)
+B^2e^{-2t}$$ \be \tau _{-t}=xe^{-\varphi} e^{-t} +O(e^{-2t}) \ee
where
$$ e^\varphi = {1\over 4} (u+v-2a)$$
\be  x={1\over 2}(u-v)\ee
$$ B= {1\over 4} (u+v+2a) $$
To obtain the limiting behaviour of the vertex operators, we will begin with
the case of $\chi \neq  0$. In this case:
\be W_\omega ^\lambda (t) \stackrel {t\rightarrow \infty} \longrightarrow
e^{-2ix\chi e^{-\varphi}} (y_{-t})^{-i\chi e^t}F(a,b,1,1-y_{-t}) \ee
where:
$$ a={1\over 2} -i\chi e^t -i\lambda$$
$$ b={1\over 2} -i\chi e^t +i\lambda$$
To study the large $z$ behaviour of the
above $F(a,b,c,z)$, we convert
it into a combination of $F(\alpha,\beta,\gamma,{1/ z})$
$^{[15]}$ ,
\be F(a,b,1,z=1-y_{-t})= B_1 (-z)^{-a} F(a,a,1-2i\lambda, {1\over z})\\
+B_2(-z)^{-b} F(b,b,1+2i\lambda,{1\over z}) \ee
where
$$B_1={\Gamma (2i\lambda)\over \Gamma (b) \Gamma (1-a)}$$
$$ B_2={\Gamma (-2i\lambda)\over \Gamma(a) \Gamma (1-b)}$$
Now in our case, $a$ and $b$ also tend to infinity
and the result is,
$$F(a,a,1-2i\lambda,{1\over z}) \rightarrow  \ \ \ _0F_1(1-2i\lambda, \chi ^2
e^{-2\varphi})$$
$$F(b,b,1+2i\lambda , {1\over z})\rightarrow  \ \ \ _0F_1 (1+2i\lambda, \chi ^2
e^{-2\varphi})$$
Then using the following identities$^{[15]}$:
$$_0F_1 (\nu +1,{z^2\over 4})=\Gamma (\nu+1)({z\over 2})^{-\nu} I_\nu (z)$$
$$\Gamma (i\chi e^t+{1\over 2}+i\lambda)=
(i\chi e^t)^{i\chi e^t} e^{-i\chi e^t} \sqrt {2\pi} (i\chi e^t)^{i\lambda} ,
$$
where $I_\nu$ is the modified Bessel function of the first kind, we can show
tha
t:
\be F(a,b,1,z=1-y_{-t}) \rightarrow {e^{2(\varphi +t)i\chi e^t}
e^{-(t+\varphi)}\over 2\pi e^{-\pi \chi e^t}}
2i\lambda \Gamma (2i\lambda) \Gamma (-2i\lambda)
[I_{2i\lambda}(2\chi e^{-\varphi})-I_{-2i\lambda}(2\chi e^{-\varphi})]\ee
But as:
$$K_\nu (z)={\pi \over 2} (\sin \nu \pi)^{-1} [I_{-\nu}(z) -I_\nu(z)]$$
$$\Gamma (z) \Gamma (-z) =-\pi z^{-1} \csc (\pi z)$$
where $K_\nu (z)$ is modified Bessel function of the third kind, we find:
$$F(a,b,1,z=1-y_{-t})={1\over \pi} e^{\pi \chi e^t} e^{2i\chi (\varphi +t)
e^t}e^{-(\varphi+t)}K_{2i\lambda} (2\chi e^{-\varphi}).$$
We will therefore find the following expression for the vertex operators in the
region V(and III):
\be W_\chi^\lambda(t) \stackrel {t\rightarrow \infty} \longrightarrow
 {1\over \pi}e^{-2ix\chi e^{-\varphi}}e^{\pi \chi e^t}
e^{-(t+\varphi)}K_{2i\lambda}(2\chi e^{-\varphi})\ee
Fortunately the dependence of $W_\chi ^\lambda$ on $t$ is such that we can
absorb it consistently in $\varphi$, and
therefore if we define $\phi=\varphi +t$ and $xe^{-\phi}=X$  and use
Eq.(9), we will finally obtain,
\be W_\omega^\lambda (t) \stackrel {t\rightarrow \infty}\longrightarrow
{e^{\pi \omega}\over \pi}e^{-2i\omega X}
e^{-\phi}K_{2i\lambda}(2 \omega e^{-\phi}) \ee
The Eq.(19) is exactly the vertex operator of c=1 coupled to 2-d gravity with
no
n-zero
cosmological constant $^{[16]}$. This equivalence is even more clear when the
vertex operator
is considered on-shell, that is when $\lambda =\pm {\omega / 3}$ $^{[3]}$.
Eq.(19) also shows that the eigenvalue of $\sigma$
plays the role of the cosmological constant  $\chi =\sqrt {\mu}$
$^{[3]}$.

To obtain the limit of $W _\omega ^\lambda (t)$ for the case $\chi
\rightarrow 0$, we go back to Eq.(16). In this case
$a$ and $b$ are finite and ${1/z}\rightarrow -e^{-2\varphi} e^{-2t}$. So
by expanding $_2F_1$'s in the right hand side of Eq.(16)
and taking only the leading term, we find:
\be W_\omega^\lambda (t) \stackrel {t\rightarrow \infty} \longrightarrow
e^{-(\varphi+t)} e^{-ix\chi e^{-\varphi}} \{Ae^{2i\lambda(\varphi +t)}+c.c.\}
\e
e
where:
$$A={\Gamma (2i\lambda) \over \Gamma ({1\over2}+i\lambda +i\omega)\Gamma
({1\over 2}+i\lambda - i\omega)}$$
As in the previous case, if we define $\phi =\varphi + t$ and
$X =xe^{-\phi}$
we will arrive at the following expression for the primary fields
\be W _\omega^\lambda \rightarrow e^{-\phi} e^{-2i\omega X } \{
Ae^{2i\lambda \phi}+A^* e^{-2i\lambda \phi}\} \ee
However, this is nothing but the vertex operator of c=1+Liouville at zero
cosmological
constant, of course after applying the on-shell condition. The
scattering matrix is ${A/ A^*}$.
Therefore $\chi $ plays exactly the role of the cosmological constant
and the boosted black hole in region V is equivalent to 2-d gravity.

At $t=\infty$ where $\chi =\omega e^{-t}=0$ (for finite value of energy),
Eq.(20) becomes:
 \be W_\omega^\lambda (t=\infty) = Ae^{2i(\lambda -1)\phi} +c.c. \ee
which is the expression for
the vertex operator of pure gravity, Liouville theory, as expected:
as $t \rightarrow \infty$, $\sigma _3^t \rightarrow \sigma$ and
we expect the theory to reduce
to $SL(2,R) / E(1) $, which is the Liouville theory.

\section{Free field representation of $SL(2,R)/ E(1)$}
In this section we will show the equivalence of $SL(2,R) / E(1) $
and Liouville theory
by looking at the stress tensor . As is well known,
if one uses the Gauss
decomposition to represent the group elements of $SL(2,R)$, the following
repres
entations
for the currents of $SL(2,R)_k$ in terms of free fields $\beta$,$\gamma$ and
$\p
hi$ can be
obtained $^{[13,17]}$:
$$ J_+ = \beta $$
\be J_- = \beta \gamma^2 +\sqrt {2k'} \gamma \partial \phi +k \partial
\gamma \ \ \ ,  \ \ \ \ k'=k-2 \ee
$$J_3= -\beta \gamma - \sqrt {k'\over 2} \partial \phi $$
where $\beta$ and $\gamma$ are the commuting ghost fields with dimensions
$h=1,0
$
and with OPE's $\beta (z) \gamma (w) \sim {1\over z-w}$ and $\phi (z)\phi (w)
\s
im -\lg (z-w)$. Then using Sugawura construction,
the stress tensor of $SL(2,R)_k$ becomes:
\be T _{SL(2,R)}(z) = \beta \partial \gamma -{1\over 2} (\partial \phi)^2
-{1\over \sqrt {2k'}} \partial ^2 \phi \ee
Now we want to gauge away the nilpotent subgroup of $SL(2,R)$,i.e. $J_+$ by
usin
g the
BRST method. As $J_+ (z) J_+ (w) $ is regular, there is no need to
introduce a gauge field
(auxiliary field) for constructing the BRST charge ($Q _+$),and hence there is
n
o
need to introduce ghosts to fix the gauge field. In this way we arrive at the
following
expression for the BRST charge of the nilpotent subgroup of $SL(2,R)$:
\be Q_+ = \oint dz J_+ (z) \ee
which satisfies,
$$Q_+ ^2=0$$
As we do not introduce the gauge field, so it has no contribution in $T(z)$ and
therefore
\be T_{SL(2,R)/ E(1)}=\beta \partial \gamma -{1\over 2} (\partial
\phi)^2 -{1\over \sqrt {2k'}} \partial ^2\phi \ee
But there are terms in Eq.(26) which are BRST exact and must be subtracted
from the stress tensor. It can be easily shown that:
$$\beta \partial \gamma =\partial (\gamma \beta) - \gamma \partial \beta$$
$$= \partial (\gamma \beta) -\{Q_+, {1\over 2} \partial \beta \gamma^2\}$$
 $$= \{Q_+, \partial ({1\over 2} \beta \gamma^2)\} - \{Q_+, {1\over 2} \partial
\beta \gamma^2\}$$
Thus up to a BRST exact term and at the level $k={9\over 4}$, we find that:
\be T_{SL(2,R)/E(1)} = -{1\over 2}(\partial \phi)^2 -\sqrt {2}
\partial ^2 \phi \ee
The above equation is exactly the Liouville action at zero cosmological
constant.The same result can be otained if we consider the stress tensor of
$SL(2,R) / U_t(1) $ and look at its behaviour at $t \rightarrow \infty $ .

{\large Acknowledgements}

We would like to thank A.Morozov for valuable discussions on the free field
representation of GWZW models.

\begin{center}
{\large  References \\}
\end{center}
\begin{enumerate}

\bibitem{1}  E.Witten, Phys. Rev. D44 (1991) 314
\bibitem{2}  G.Mandel, A.M.Sengupta and S.R.Wadia, Mod. Phys. Lett. A6(1991)
168
\bibitem{3}  R.Dijkgraaf, H.Verlinde and E.Verlinde, Nucl.Phys.B 371 (1992) 269
\bibitem{4}  A.A.Tseytlin,Nucl.Phys. B399 (1993)601\\I.Bars, Phys.Lett.B293
(1992)315
\bibitem{}  E.J.Martinec and S.L.Shatashvili,Nucl.Phys. B368 (1992) 338
\bibitem{}  J.Distler and P.Nelson, $ibid$ 374(1992) 123
\bibitem{}  S.Mukhi and C.Vafa, $ibid$ 407 (1993)667
\bibitem{}  T.Eguchi, H.Kanno and S.K.Yang,Phys.Lett. B298 (1993)73
\bibitem{}  T.Eguchi, $ibid$ 316(1993) 74
\bibitem{}  M.Ishikawa and M.Kato, $ibid$ 302 (1993) 209
\bibitem{}  S.Chaudhuri and J.D.Lykken ,Nucl.Phys. B396 (1993) 270 \\ K.Becker
            and M.Becker : Interactions in the $SL(2,R)/U(1)$ blackhole
            background , CERN-TH.6976/93
\bibitem{}  N.Marcus and Y.Oz, hepth/9305003
\bibitem{}  M.Bershadsky and D.Kutasov, Phys. Lett. B266 (1991) 345
\bibitem{}  M.Alimohammadi,F.Ardalan and H.Arfaei, Int. J. Mod.
Phys. A, in press\\
 F.Ardalan, Proc. of Workshop on Low Dimensional Topology and Quantum Field
Theo
ry,
Isaac Newton Inst., Sept. 1992.
\bibitem{}  Erdelyr,Magnus,Oberhettinger and Tricomi: Higher Transcendental
Functions (Bateman Manuscript project), McGraw-Hill (1953)
\bibitem{}  G.Moore, N.Seiberg and M.Staudacher: From Loops to States in 2D
Quantum Gravity, Rutgers preprint RU-91-11 (March,1991)
\bibitem{}  A.Gerasimov, A.Morozov and M.Olshanetsky, Int.Jour. of Mod. Phys.A5
(1990),2495
\bibitem{}  K.Itoh, H.Kunitomo,N.Ohta and M.Sakaguchi; BRST Analysis of
Physical States in Two-Dimensional Blackhole, OS-GE 28-93

{\large Figure Caption } \\
Fig. 1:The two-dimensional black hole in $(u,v)$-coordinates.

\end{thebibliography}

\end{document}